\documentclass[fleqn,10pt]{wlscirep}
\usepackage[utf8]{inputenc}
\usepackage[T1]{fontenc}

\title{Start-to-end modelling of laser-plasma acceleration, beam transport and dose deposition of very high-energy electrons for radiotherapy}

\author[1,2]{Rajakrishna Kalvala}
\author[1,2,*]{Anton Golovanov}
\author[1]{Arnaud Courvoisier}
\author[1]{Tomer Friling}
\author[1]{Eyal Kroupp}
\author[1]{Lidan Grishko}
\author[1]{Victor Malka}
\affil[1]{Weizmann Institute of Science, Department of Physics of Complex Systems, Rehovot, 7610001, Israel}
\affil[2]{These authors contributed equally to this work}
\affil[*]{Corresponding author: anton.golovanov@weizmann.ac.il}

\usepackage{siunitx}
\usepackage{physics}

\begin{abstract}
Radiotherapy using very high-energy electron (VHEE) beams generated by a laser-plasma accelerator has garnered significant interest due to its dose distribution capabilities and potential to address limitations of traditional photon-based radiotherapy. To explore the feasibility of such approach, the presented study uses the parameters of OONA, the commercial <1.3\,J, <25\,fs pulse duration laser recently installed at the Weizmann Institute of Science. Through particle-in-cell simulations of laser-plasma interaction, realistic electron beams were obtained. After filtering and collimation with a beamline and arranging them into an array of pencil beams, their radiotherapeutic potential was investigated. GEANT4 simulations were used to calculate the dose deposition in water phantoms and heterogeneous phantoms with bone inserts, considering realistic beam parameters for laser-plasma accelerators. Multifield irradiation setup and the dose distribution at the isocenter through different incidence angles were studied, simulating an intensity-modulated delivery. Our findings demonstrate that polychromatic VHEE beams generated from laser-plasma accelerators, when collimated using a compact beamline of quadrupoles and dipoles, can deliver an on-axis dose with enhanced precision and uniformity. This study highlights the transformative potential of laser-plasma accelerators in advancing radiotherapy modalities, paving the way for further research and clinical implementation.
\end{abstract}
\begin{document}

\flushbottom
\maketitle
\thispagestyle{empty}


\section*{Introduction}

Radiotherapy remains a cornerstone of cancer treatment, with approximately 50\% of all cancer patients receiving some form of radiation therapy during their treatment course. \cite{delaney2005role, delaney2015evidence} Conventional photon radiotherapy remains limited in its ability to effectively treat deep-seated tumours while adequately sparing surrounding healthy tissues. Photon beams result in unavoidable entrance and exit dose deposition beyond the target volume, constraining treatment optimization. \cite{baumann2016radiation} In comparison, proton therapy offers superior dose conformity through its characteristic Bragg peak, but its widespread implementation is hindered by high infrastructure costs and accessibility challenges, with fewer than 100 centres worldwide serving the global cancer population. \cite{paganetti2006proton, song2024review} This substantial gap between treatment availability and clinical needs \cite{rosenblatt2018radiotherapy, zubizarreta2017analysis} underscores the urgent necessity for innovative radiotherapy modalities that can combine the dosimetric advantages of particle therapy with the accessibility and cost-effectiveness of conventional systems.

To improve the effectiveness of radiation oncology and enhance patients' health-related quality of life, radiotherapy must be capable of inducing cancer cell death while preserving healthy tissues. \cite{baumann2016radiation} Current studies highlight very-high energy electron (VHEE) radiotherapy as a promising approach to achieve these goals. \cite{bohlen20233d, zhang2023treatment, kokurewicz2019focused, svendsen2021focused, guo2025preclinical, zhou2025compactdosedeliverylaseraccelerated}
VHEE in the 150--250\,MeV range were first proposed by DesRosiers et al. as a novel radiotherapy modality allowing the treatment of deep-seated tumours.\cite{desrosiers2000150} Since then, several studies have emphasized the advantages of VHEE use in radiotherapy. Unlike conventional electron therapy, which is typically limited to treating superficial tumours due to the low penetration depth of 5--25 MeV electrons, VHEE beams can effectively reach deep-seated tumours at depths of 10--15\,cm with millimeter precision. \cite{bazalova2015treatment, labate2020toward} Combined with the insensitivity of VHEE treatment to the anatomical heterogeneities, it represents a significant advancement in radiation oncology, potentially expanding the therapeutic options available for treating complex and challenging malignancies. \cite{schuler2017very, ronga2021back, lagzda2020influence, lagzda2017relative} Moreover, to increase the conformal dose delivery to deep targets, VHEE beams of more than 100 MeV were found to be desirable for intensity-modulated radiotherapy. \cite{yeboah2002optimization} Comparative treatment planning studies between conventional radiotherapy modalities and VHEE have shown lower spinal cord dose in the pediatric brain case and lower brain stem dose in the VHEE case. \cite{schuler2017very, palma2016assessment, bazalova2015treatment}
Additionally, VHEE beams can be electromagnetically scanned or focused with high precision \cite{kokurewicz2019focused, an2024optimizing}, enabling more accurate dose delivery to the tumour volume while minimising exposure to surrounding healthy tissues. \cite{whitmore2021focused, bedford2024treatment, svendsen2021focused}

VHEE beams have also started gaining wider interest with the advent of laser-plasma accelerators (LPAs) that can generate ultra-high accelerating fields, significantly reducing the size and cost of accelerator facilities compared to conventional linear accelerators. \cite{tajima1979laser, malka2008principles, esarey2009physics, malka2012laser}
Recent advancements in LPA technology have significantly improved electron beam quality and succeeded in achieving low energy spread and enhanced shot-to-shot stability in the energy range suitable for VHEE radiotherapy \cite{faure2006, kirchen2021optimal}.
Even though in several aspects the quality of the beams remains below the level of conventional accelerators, these advancements make LPAs favorable candidates for future clinical applications. \cite{glinec2006radiotherapy, desrosiers2008laser, Malka_2010_MR_704_142, ronga2021back, panaino2025very, guo2025preclinical}
Unique properties of LPA-generated electron beams compared to beams from conventional linacs require specific beamline designs\cite{svendsen2021focused, whitmore2021focused, zhou2025compactdosedeliverylaseraccelerated} and paying attention to the influence of the energy spread and angular divergence on dose delivery.
This poses an additional challenge for numerical simulations, as a rigorous investigation must include generation of an electron beam in a LPA, its propagation through the beamline, and the following interaction with the phantom.
However, previous simulations of LPA-based VHEE used approximate models of LPA-generated beams instead.\cite{glinec2006radiotherapy,lundh2012comparison,labate2020toward}
Additional challenges include the development of appropriate dosimetry techniques for accurately measuring and monitoring VHEE beams, managing secondary radiation (particularly neutron production), and designing compact, cost-effective systems suitable for clinical environments. \cite{ronga2021back, subiel2014dosimetry}

In this study, we employ a start-to-end computational approach to evaluate the potential of laser-plasma acceleration (LPA) for generating very high-energy electron (VHEE) beams in radiotherapy. Particle-in-cell (PIC) simulations are used to model the interaction between an intense laser pulse and plasma, optimizing the conditions for producing suitable electron beams.
Simulations are performed for different gas target parameters with different target energies of the accelerated beams.
The generated beam is then processed through beamline simulations, including filtering and collimation, to refine its energy spectrum and spatial characteristics for therapeutic applications.
Finally, GEANT4 simulations using these PIC generated beams are conducted to analyze dose deposition patterns in both water and heterogeneous phantoms.
In these simulations, we consider an interaction geometry in which a \numproduct{7x7} array of LPA-produced beams steered by dipole magnets converges at the same spatial point.
This geometry is shown to reduce the entrance dose and increase the longitudinal localization of the dose distribution.
The possibility of using this array configuration in a multi-field irradiation configuration is also studied.
This start-to-end approach enables the study of LPA-generated VHEE beams for precise and effective cancer treatment and paves the way towards integrating such beams into clinical treatment planning systems.

\section*{Results}
The numerical study of dose deposition in phantoms is critical for optimizing radiation therapy techniques, as accurate simulations help to predict dose distributions and enhance treatment efficacy. Here we investigate how different electron beam parameters that are generated in laser-plasma accelerators and propagated through a beamline affect the 3D dose deposition in a water-equivalent phantom and a heterogenous phantom with a bone insert of 2\,cm using PIC and GEANT4 simulations.

\subsection*{LPA stage}

To simulate the generation of an electron beam in a LPA, we perform PIC simulations of the interaction of a laser pulse with a gas jet using FBPIC. \cite{lehe2016spectral}
In the simulations, a 25\,fs (intensity FWHM) laser pulse with energies of either 0.3\,J or 0.5\,J is interacting with a jet of air or helium--nitrogen mixture.
Such a short and high-power laser pulse almost fully ionizes the gas and drives a nonlinear wakefield in the resulting underdense plasma.
The electrons are then injected into the plasma wakefield through ionization injection, a process when electrons from lower shells of nitrogen and oxygen are ionized near the peak field of the laser pulse already inside the wakefield which ensures their trapping.\cite{Pak_2010_PRL_104_25003}
In the case of the helium--nitrogen mixture, helium serves as the background gas, and the amount of the injected electrons can be controlled by changing the concentration of nitrogen.
Air always provides very strong ionization injection as both of the main gas components (nitrogen and oxygen) have lower shells which are ionized only inside the laser pulse at the considered intensity.
The parameters of the gas jet are chosen to produce electron bunches with a high charge and target energies of 200\,MeV, 250\,MeV, and 300\,MeV.

\begin{figure}[tb]
    \centering
    \includegraphics[width=\linewidth]{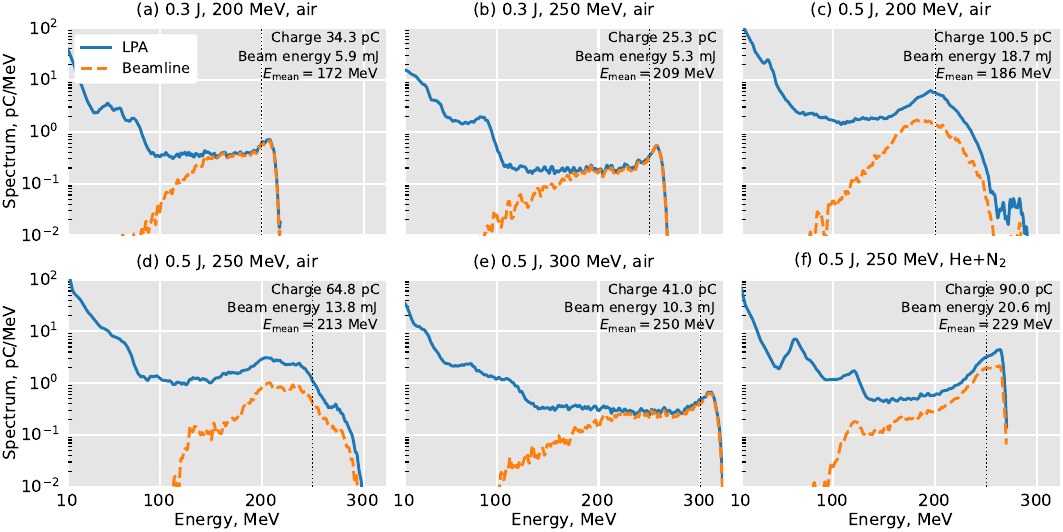}
    \caption{Energy spectra of electron beams after the laser--plasma accelerator (solid lines) and after the beamline (dashed lines) for different laser energies (0.3\,J and 0.5\,J), different target energies (200\,MeV, 250\,MeV, 300\,MeV) of the electron beam, and different gas mixtures (air or a helium--nitrogen mixture).
    The total charge, the total energy and the mean electron energy $E_\mathrm{mean}$ for the beam after the beamline are specified for each case.
    The vertical dotted lines show the target electron energy.}
    \label{fig:spectra}
\end{figure}

The resulting spectra from PIC simulations are shown with solid lines in Fig.~\ref{fig:spectra}.
In all cases, we observe electron bunches with a wide energy spectrum with a high number of low-energy electrons, which is typical for the ionization injection scheme.
However, the spectra have noticeable peaks near the target energy.

\subsection*{Beam transport line}

Immediately at the exit from the LPA stage, the electron beam has a very small transverse size (several \si{\um}) but a relatively high divergence (typically several mrad).
To improve the electron beam spectrum by filtering out low-energy electrons as well as to collimate the beam, we utilize a beam transport line.

\begin{figure}[tb!]
    \centering
    \includegraphics[width=1\linewidth]{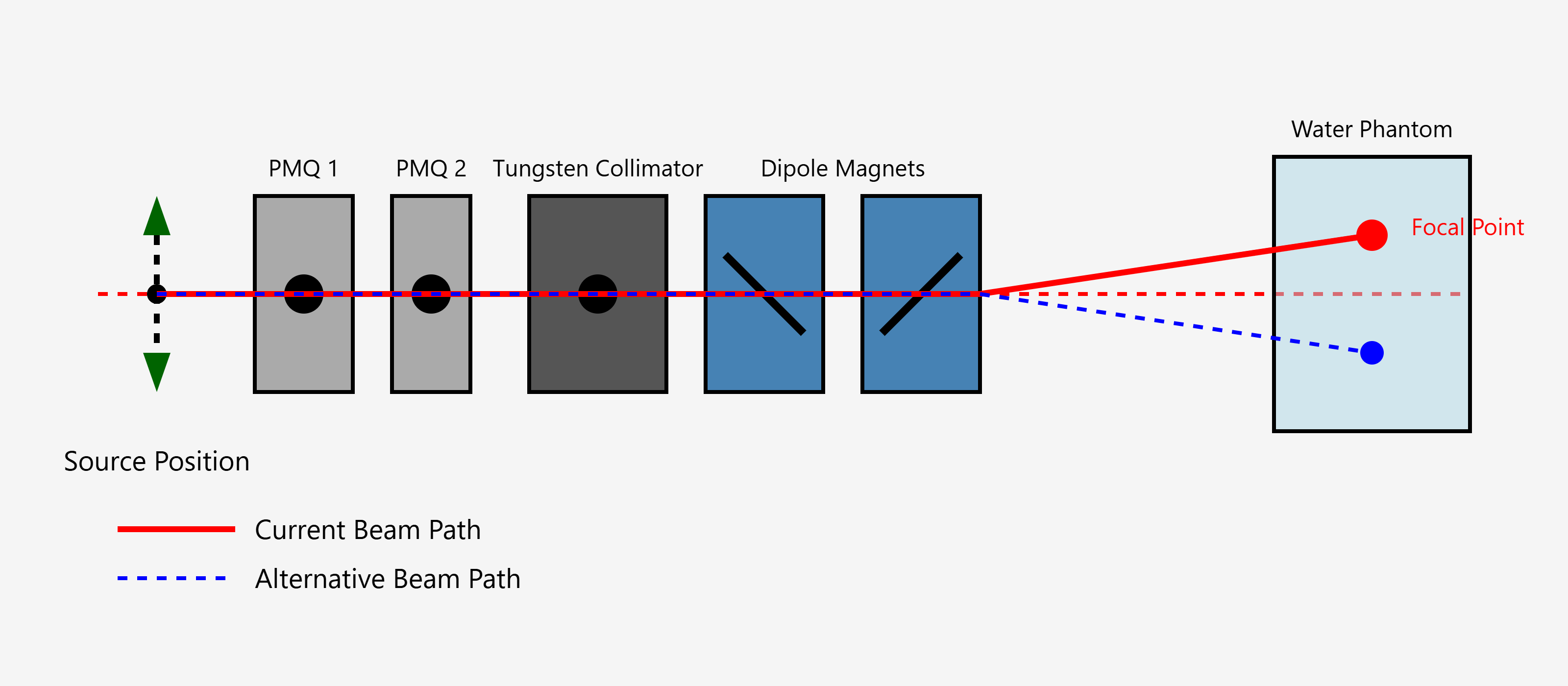}
    \caption{A schematic of the beamline with two permanent magnet quadrupoles (PMQs), a collimator, and two dipole magnets.}
    \label{fig:ebeam4therapysystem}
\end{figure}

The design of the beam transport line from the LPA stage to the phantom is illustrated in Fig.~\ref{fig:ebeam4therapysystem}. The beam transport line consists of two permanent magnet quadrupoles (PMQs) for the collimation of the electron beam,\cite{fuchs2009laser, pompili2018compact} a collimator to prevent highly divergent low-energy electrons from reaching the phantom, and two dipole magnets to steer the beam to the desired angle in two perpendicular directions.
The quadrupoles have a focusing gradient of 545\,T/m, lengths of 2\,cm and 1.5\,cm, respectively, and an inner radius of 3\,mm.
They provide focussing in two different perpendicular directions, respectively, with the first quadrupole providing focussing in the laser polarization direction.
Their position is adjusted to ensure proper collimation of electrons at the target energy and therefore changes for different target energies.
The collimator is located 15\,cm away from the center of the second quadrupole and has a pinhole with a radius of 1\,mm.
The dipoles have a length of 1.5\,cm and an adjustable magnetic field of up to $\pm\SI{2}{T}$ each.

The beamline is designed for a particular energy, and any electrons that are farther away from the designed energy will be scatter by the quadrupoles and consecutively filtered out by the collimator device positioned after the quadrupoles.
Lower energy electrons from LPAs typically experience larger divergence and can scatter even before reaching quadrupoles.
Dashed lines in Fig.~\ref{fig:spectra} show the energy spectra of electron beams after the collimator.
In all cases, we observe a very significant reduction in the low-energy part.
In some cases, the reduction happens also for the target energy, owing to high divergence of a beam from the LPA stage which results in the collimated beam size being larger than the pinhole radius.

\begin{figure}[tb!]
    \centering
    \includegraphics[width=\linewidth]{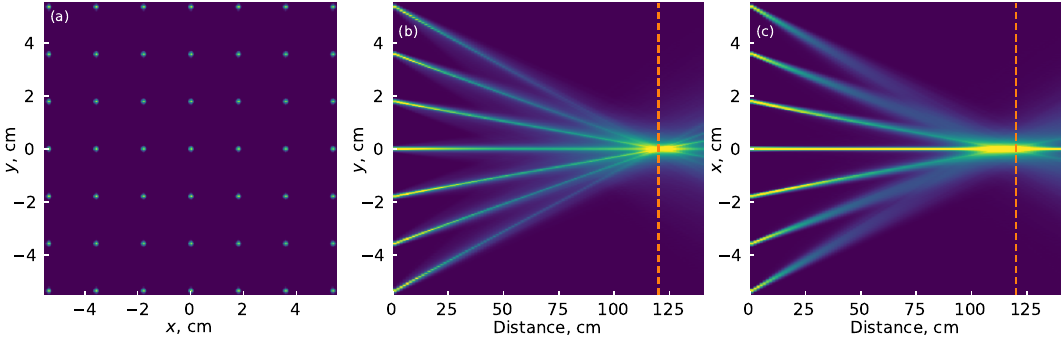}
    \caption{The initial transverse electron density distribution (a) and the evolution of the electron density projections to the $y$ (b) and $x$ (c) directions with the propagation distance for the propagation of a $7 \times 7$ array of beams in vacuum.
    The used beam corresponded to Fig.~\ref{fig:spectra}(c).
    The vertical dashed lines show the expected conversion point (in the absence of the phantom) 1.2\,m away from the beamline's exit.}
    \label{fig:grid_focussing}
\end{figure}

The steering capability of the two dipoles located after the collimator is used to form a $7 \times 7$ array of beams converging at the same point (Fig.~\ref{fig:grid_focussing}).
In practice, it can be achieved either by moving the source and the beamline assembly and or by moving the phantom in the transverse direction after each laser shot (or several consecutive shots with the same angle), while adjusting the magnetic field of the dipoles to ensure they pass through the same point in the phantom.
The required magnetic fields of the dipoles are calculated for the target bunch energy.
Because the bunches after the collimator still have a relatively wide spectrum, steering for larger angles also introduces additional divergence of the beam, as the deflection angle is inversely proportional to the electron energy.
When using the array, this leads to the extended focus in the longitudinal direction.

\subsection*{3D Dose deposition calculation}

The propagation characteristics of polychromatic electron beams generated by LPA within a water-equivalent phantom are investigated by using realistic electron bunches generated from PIC simulations and filtered by the beamline. To validate the proposed design model, 3D dose deposition calculations are conducted using Monte Carlo--based GEANT4 simulations.
The edge of the phantom is located 1\,m away from the end of the beamline, and it is assumed that the space between the beamline and the phantom is filled with air.

\subsubsection*{Dose Distribution in Homogeneous Phantom}

\begin{figure}[tb!]
\centering
\includegraphics[width=\linewidth]{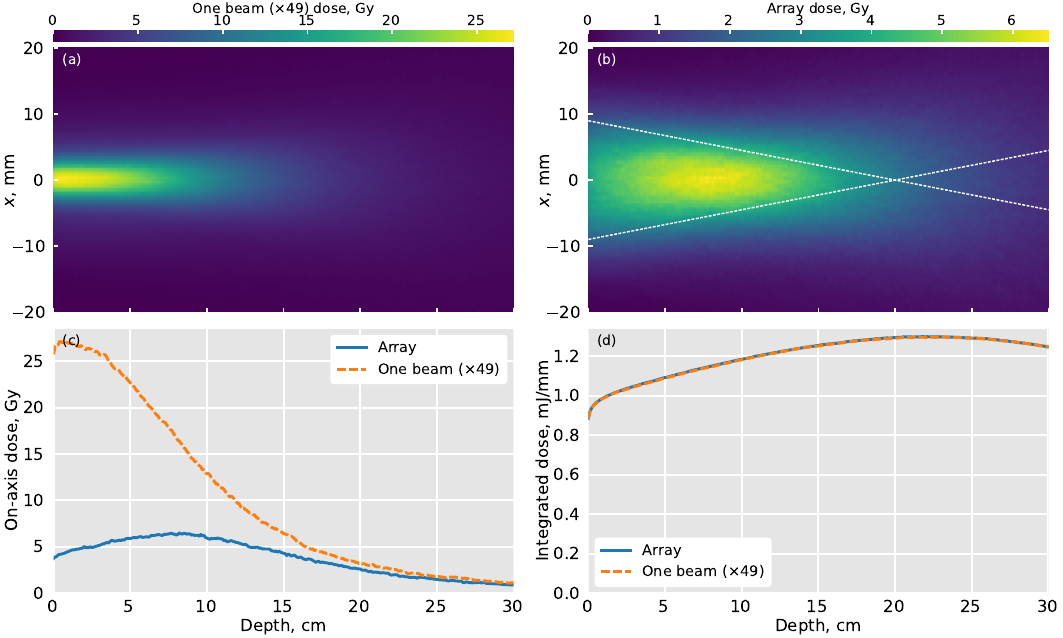}
\caption{Dose distribution in the \textit{zx} plane (\textit{y} = 0) of the water phantom for the cases of (a) using one beam and (b) an array of beams for the beam corresponding to Fig.~\ref{fig:spectra}(c). The dashed lines in (b) show the trajectories of 200 MeV electrons from the beams with the maximum incidence angle in the absence of the phantom. The dependences of the (c) on-axis dose (\textit{x} = \textit{y} = 0) as well as the integrated over the transverse \textit{xy} plane dose. }
\label{fig:Dose Distribution using array of beams}
\end{figure}

As described in the previous section, we propose using an array of beams incident at different angles to create a more optimal dose distribution. The dose distribution was analyzed through the calculation of the on-axis dose and integrated dose profiles. The on-axis dose provides a measure of the maximum dose delivered along the central axis of the beams, while the integrated dose evaluates the dose deposition as a function of depth within the phantom. Fig.~\ref{fig:Dose Distribution using array of beams} shows the comparison between the dose deposition created by a $7 \times 7$ array of beams (see Fig.~\ref{fig:grid_focussing}) incident at different angles in both \textit{x} and \textit{y} directions to one normally incident beam (repeated 49 times to achieve the same total dose absorbed by the phantom). The dose distribution from one collimated beam is very narrow around its entrance to the phantom. However, it quickly expands in the transverse direction with depth due to the scattering of electrons inside the phantom, leading to a drop in the on-axis dose. This distribution is suboptimal because of a much larger entrance dose compared to the in-depth dose. However, when we use an array of beams incident at different angles, the peak on-axis dose moves deep inside the phantom, although it does not reach the same depth as the intended convergence point of all the beams in vacuum due to both the scattering and the dependence of the convergence point coordinate on the electron energy. At the same time, the distribution of the dose integrated over the transverse plane does not depend on the interaction geometry, as the incidence angles of the beams remain very small.

\begin{figure}[tb!]
\centering
\includegraphics[width=1\linewidth]{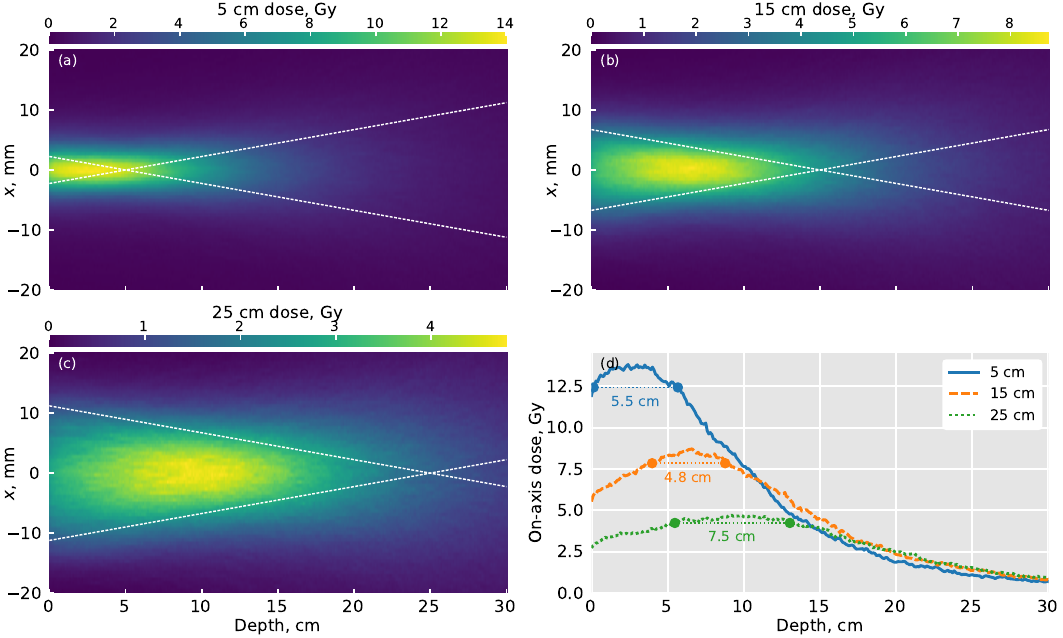}
\caption{Dose distribution in the \textit{zx} plane (\textit{y} = 0) of the water phantom for different depths of intended beam array convergence (in vacuum): (a) 5 cm, (b) 15 cm, (c) 25 cm for the 200\,MeV beam corresponding to Fig.~\ref{fig:spectra}(c). The dashed lines show the trajectories of 200 MeV electrons from the beams with the maximum incidence angle. (d) The dependencies of the on-axis (\textit{x} = \textit{y} = 0) dose on the depth for all three cases. The horizontal lines with values show the therapeutic range (width at 90\% peak dose level).
}
\label{fig:Dose distribution in different depths}
\end{figure}

By moving the point of convergence of the beams inside the phantom, we can achieve a different depth of the peak dose deposition (see Fig.~\ref{fig:Dose distribution in different depths}).
The figure demonstrates that for VHEE therapy with the considered array geometry, the entrance dose is significantly lower relative to the peak dose delivered.
Choosing the position of the convergence point provides flexibility in targeting tumours located at different depths while reducing the entrance dose.
The ability to steer the beam with compact dipole magnets to generate array distributions is unique to electrons and constitutes an advantage over conventional photon and proton techniques.

\begin{figure}[tb!]
\centering
\includegraphics[width=1\linewidth]{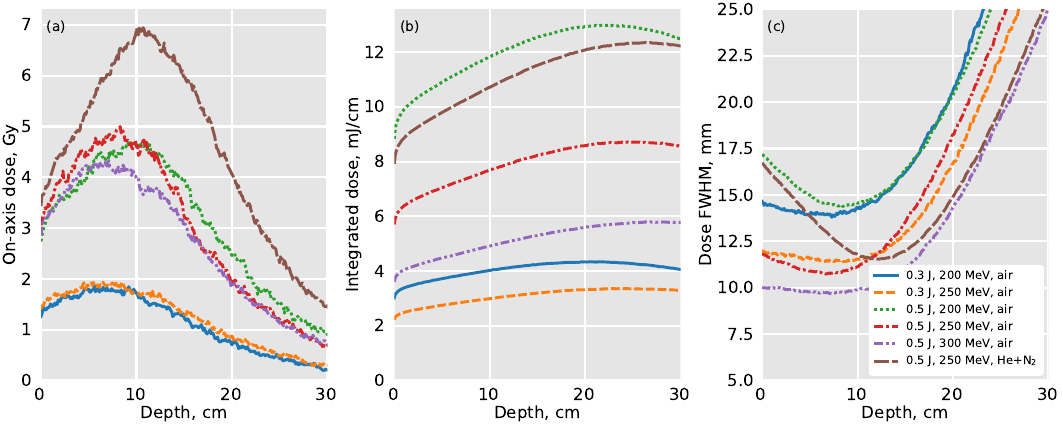}
\caption{The dependencies of the (a) on-axis dose (\textit{x} = \textit{y} = 0), (b) the integrated over the transverse \textit{xy} plane dose, and (c) the transverse size (FWHM) of the dose distribution for all beams from Fig.~\ref{fig:spectra} when using a $7\times7$ array with expected convergence (in vacuum) at the depth of 25 cm. }
\label{fig:The dependencies on-axis dose}
\end{figure}

The comparison of the distributed dose properties when using arrays of different realistic beams from PIC simulations are shown in Fig.~\ref{fig:The dependencies on-axis dose}. In all cases, the dose distribution experiences a similar behavior of having a peak deep inside the phantom. The deepest dose peak is reached by a 200\,MeV beam generated by a 0.5\,J laser, as well as for a 250\,MeV beam generated in a helium--nitrogen mixture, which can be explained by their comparatively lower relative energy spreads.

\subsubsection*{Dose Distribution from Multi-Field Irradiation}

\begin{figure}[tb!]
    \centering
    \includegraphics[width=1\linewidth]{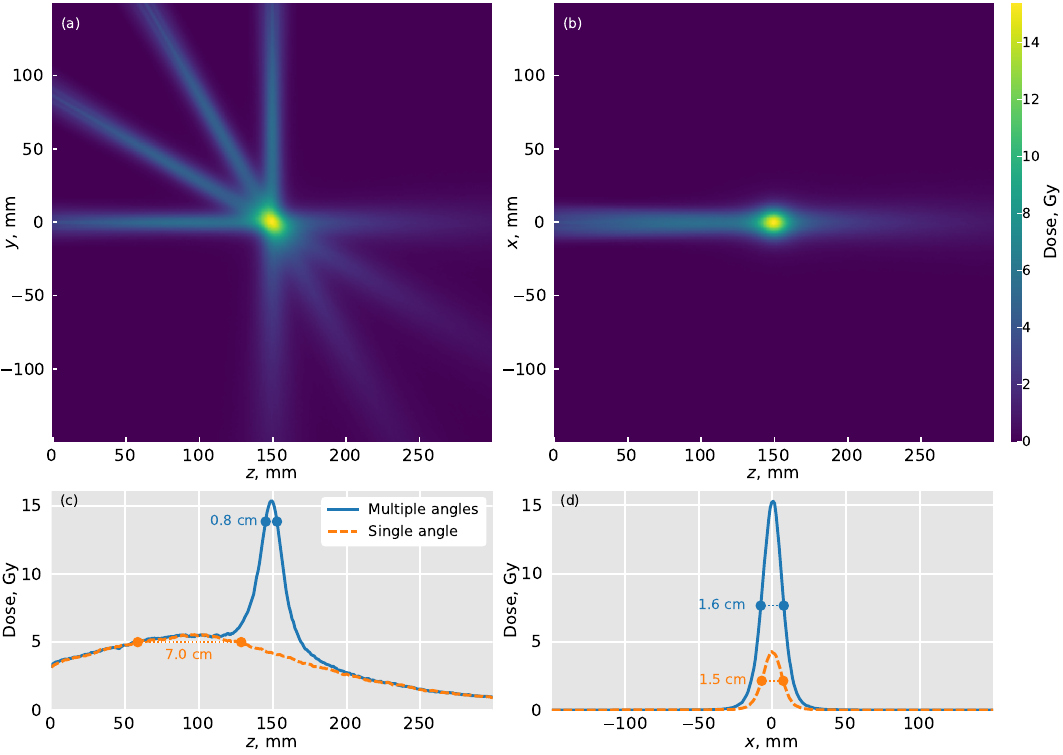}
    \caption{Multiple field irradiation in 0°, 30°, 60°, and 90° angles at the isocentre of the phantom. (a) and (b) show the dose distribution in the $zy$ and $zx$ planes; (c) and (d) show the distribution along the $z$ and $x$ axes.
    For comparison, the dose distribution from a single array propagating along $z$ is shown with a dashed line in (c) and (d).
    The horizontal lines with values in (c) show the therapeutic range (width at 90\% peak dose level), in (d) FWHM of the distribution.}
    \label{fig:multifieldirradiation}
\end{figure}

Four arrays of 49 collimated very high-energy electron (VHEE) beams were directed toward the isocenter \cite{gardner1972tumor} with focal point of 15 cm from the surface of the phantom measuring \qtyproduct{30x30x30}{cm}. To achieve comprehensive dose coverage, the beam arrays were delivered at incidence angles of 0°, 30°, 60°, and 90°. This multi-angle arrangement ensured the increased localization of the dose distribution and the reduction of the relative dose received by the surrounding volume which may represent healthy tissues. Each beam delivered a fraction of the total dose, and the angular distribution was arranged to conform the dose to the isocenter and target volume. In total, 196 beams converged at the isocenter, to evaluate combined dose deposition profile. The used beams corresponded to Fig.~\ref{fig:spectra}(c) with the expected convergence in vacuum at the depth of 25\,cm (similar to the 25\,cm case in Fig.~\ref{fig:Dose distribution in different depths}).

Figure~\ref{fig:multifieldirradiation} demonstrates the dose distribution in the multifield irradiation setup.
The therapeutic volume (the volume in which more than 90\% of the peak dose is delivered) \cite{radiationonclogy} is \SI{189}{mm^3}, while the therapeutic range (the depth at which more than 90\% of the peak dose is delivered) is 8 mm. The peak dose delivered at the centre of the phantom is around 15 Gy. The FWHM calculated in $x$, $y$ and $z$ directions are 16 mm, 26 mm and 26 mm, respectively.

This method of conformal radiotherapy can effectively target tumour tissues with high precision.  This study underscores the feasibility and versatility of compact beamline configurations for laser-plasma accelerator-based radiotherapy systems, paving the way for advanced radiotherapy conditions in cancer treatment.

\subsubsection*{Dose Distribution in Heterogeneous Phantom}

The effect of tissue heterogeneity on dose distribution calculations presents a significant challenge in radiotherapy planning \cite{avanzo2020electron, lagzda2020influence}. Variations in tissue density can substantially alter radiation beam paths and energy deposition patterns, particularly in proton-based radiotherapy techniques.

\begin{figure}[tb!]
    \centering
    \includegraphics[width=1\linewidth]{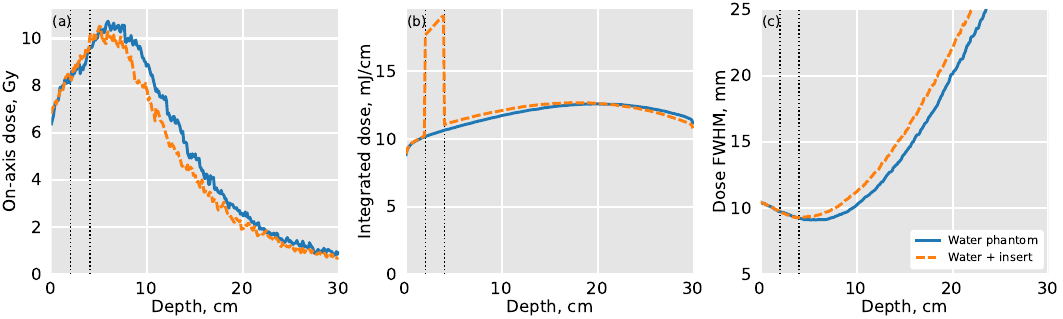}
    \caption{Comparison of dose distribution inside the water phantom and the heterogeneous phantom with a $7 \times 7$ array of beams corresponding to Fig.~\ref{fig:spectra}(c).
    The vertical lines show the boundaries of the 2-cm-thick bone insert.}
\label{fig: heterogeneous phantom dose distribution}
\end{figure}

The dose distribution in heterogeneous phantom with a 2-cm-thick bone insert with a center located at 3\,cm from the surface of the phantom was measured using the 200\,MeV beam generated by a 0.5\,J laser pulse. The resulting distributions are presented in the accompanying Fig.~\ref{fig: heterogeneous phantom dose distribution}. Our analysis of both the integrated dose distribution and on-axis dose measurements demonstrates minimal perturbation in the dose distribution pattern behind the bone insert.

This finding shows that using VHEE beams might not require complex correction algorithms to account for heterogeneities. The reduced sensitivity to heterogeneity observed in our measurements suggests potentially improved dose delivery accuracy in anatomical regions containing tissues of varying densities.

\section*{Discussion and Conclusions}

Based on the PIC simulations with FBPIC and the GEANT4 simulation toolkit, we conducted start-to-end simulations of LPA-produced VHEE beam dose deposition for radiotherapy.
Realistic beams were produced in a model of a gas jet simulated with FBPIC and propagated through a numerical model of the beamline based on a quadrupole doublet for filtering and collimating the beam.
Their interaction with a homogeneous water phantom and a water phantom with a 2 cm bone insert was then calculated with GEANT4.

Numerical measurements of 3D dose deposition were performed for an array of beams focusing at different incidence angles. The simulations show that using such an array can effectively move the peak of the dose distribution deep inside the target, thus making it possible to treat deep-seated tumours. The increased depth appears to be correlated with the reduction of the energy spread. Therefore, optimizing for a narrower energy spread while keeping the high total charge will be an important experimental task.

The use of multi-field irradiation with several arrays of beams coming from different angles around the target further demonstrated the ability to localize the peak dose and prevent damage to surrounding tissues.
The dose deposition in heterogeneous phantoms with a 2 cm bone insert showed low sensitivity of the dose distribution to inhomogeneities, possibly indicating that complex correction algorithms might not be required for effective treatment.

The results confirm that LPA-produced VHEE beams as a promising alternative to conventional radiotherapy techniques, particularly for treating tumours in challenging locations or in sensitive patient populations.

\section*{Methods}

\subsection*{PIC simulations}

We perform PIC simulations using angular mode decomposition with FBPIC \cite{lehe2016spectral}.
The laser pulse with the central wavelength of \SI{800}{nm} has a Gaussian temporal and transverse profiles at the focus, the duration of 25\,fs (power FWHM) and is focussed to a spot size of \SI{12}{\um} (radius at $1/e^2$ intensity).
Two pulse energies of \SI{0.3}{J} and \SI{0.5}{J} are considered.
These parameters correspond to the capabilities of the new OONA laser system at the Weizmann Institute of Science.

The gas jet is modeled by a trapezoidal profile with 0.2-mm-long up- and down-ramps to simulate the transition to vacuum and variable length of the plateau or variable density.
The gas composition consists of 80\% nitrogen and 20\% oxygen to model air.
As simulations rely on ionization injection, they consider nitrogen and oxygen preionized only to levels 4 and 5, respectively.

\begin{table}[tb!]
    \centering
    \begin{tabular}{lrrr}
        \toprule
        Case & Density, cm$^{-3}$ & Plateau length, mm & Laser focus, mm \\
        \midrule
        (a) 0.3 J, 200 MeV, air & $8.70 \times 10^{18}$ & 2.15 & 2.02 \\
        (b) 0.3 J, 250 MeV, air & $6.61 \times 10^{18}$ & 2.40 & 1.34 \\
        (c) 0.5 J, 200 MeV, air & $7.64 \times 10^{18}$ & 2.47 & 0.62 \\
        (d) 0.5 J, 250 MeV, air & $7.27 \times 10^{18}$ & 2.52 & 0.27 \\
        (e) 0.5 J, 300 MeV, air & $5.93 \times 10^{18}$ & 3.32 & 2.56 \\
        (f) 0.5 J, 250 MeV, He+N$_2$ & $7.67 \times 10^{18}$ & 2.30 & 2.25 \\
        \bottomrule
    \end{tabular}

    \caption{Optimized PIC simulation parameters for the beams shown in Fig.~\ref{fig:spectra}.
    The density value corresponds to the plasma density assuming partial ionization of nitrogen and oxygen to levels 5 and 6, respectively, and full ionization of helium.}
    \label{tab:beam_parameters}
\end{table}

The plasma density, the length of the plateau, and the longitudinal position of the focus are considered free parameters and are given to a genetic algorithm based on PyGAD \cite{Gad_2023_MTA} which optimizes these three parameters to provide the electron beam with the highest possible charge in the desirable energy range.
The optimized parameters are shown in Table~\ref{tab:beam_parameters}.
One of the simulations was performed with a helium--nitrogen mixture instead of air.
In this case, helium was assumed to be fully preionized, and the share of nitrogen was added as an optimization parameter. The optimal mixture was found to be 95\% helium and 5\% nitrogen (N$_2$).

In the simulations, we used a simulation box with step sizes of \SI{0.02}{\um} and \SI{0.32}{\um} in the longitudinal and transverse directions, respectively, with 3 azimuthal modes, and the time step $ct$ equal to the longitudinal step size.
Preionized atoms and electrons were initialized with $32 = 2 \text{\,(long.)} \times 2\text{\,(trans.)}\times 8\text{\,(azim.)}$ particles per cell.
To accelerate the simulations, the Lorentz boost technique was used.\cite{Lehe_2016_PRE_94_53305}

\subsection*{Beamline simulations}

All electrons with energies above \SI{5}{MeV} from the final iteration of the FBPIC output are passed to the simulated beamline corresponding to Fig.~\ref{fig:ebeam4therapysystem}.
Propagation of the beam through quadrupoles is modelled by Wake-T\cite{FerranPousa_2019_JPCS_1350_12056} simulations, assuming the quadrupoles have a uniform longitudinal distribution of the magnetic field.
The quadrupoles have an inner radius of \SI{3}{mm}, and all electrons which exceed this radius both at the entrance and the exit of the quadrupoles are filtered out.
Propagation in vacuum sections is calculated analytically (space charge effects are neglected).
The collimator is simulated by filtering out all electrons with the distance from the beamline axis above \SI{1}{mm} at its coordinate.
The magnetic dipoles are simulated by calculating analytical solutions of electron motion in the relativistic limit $\gamma \gg 1$ assuming constant magnetic field inside the dipoles.

\subsection*{Quadrupole position optimization}

Magnetic quadrupoles are electron beam transport line elements which provide focusing in one axis while defocusing in the other.
The magnetic field of a quadrupole is given by
\begin{equation}
    \vb{B} = G \gradient {(x y)},
\end{equation}
where $G = \SI{545}{T/m}$ is the magnetic field gradient.

The focal length of a doublet quadrupole lens is given by
\begin{equation}
    f = \frac{\gamma mc^2}{ec}  \frac{1}{\int G dz} = \frac{E_\mathrm{kin}}{ec} \frac{1}{GL}
\end{equation}
where $L$ is the quadrupole length, $E_\mathrm{kin}$ is the electron energy, $e$ is the elementary charge, and $c$ is the speed of light.
The imaging condition for a lens with focal length $f$ relates the source position $s$ and image position $u$ as:
\begin{equation}
    \frac{1}{f} = \frac{1}{s} + \frac{1}{u}, \quad u=\frac{sf}{s-f}
\end{equation}

When optimized, a quadrupole doublet can achieve beam collimation with the second lens positioned at a specific distance from the first lens's imaging position.The optimal configuration occurs when $z_1 = 2f_1$, $f_1 = (4/3)f_2$, and $z_2 = (8/3)f_1$, resulting in a beam size of $\sigma_x = (4/3) f_1 \theta_x$, $\sigma_y = 4 f_1 \theta_y$ after collimation, where $\theta_x$ and $\theta_y$ are initial angles of divergence.
The beam from the LPA stage usually is more divergent in the laser polarization direction, so this direction is chosen to correspond to the $x$ axis so that $\theta_x > \theta_y$.
In this case, the resulting collimated beam will have a more circular shape than before collimation.

\begin{table}[tb!]
    \centering
    \begin{tabular}{lrrrrr}
    \toprule
        Case & $f_1$, cm & $f_2$, cm & $z_1$, cm & $z_2$, cm & $z_\mathrm{coll}$, cm \\
        \midrule
        (a) 0.3 J, 200 MeV, air & 6.14 & 8.19 & 12.28 & 16.37 & 31.37 \\
        (b) 0.3 J, 250 MeV, air & 7.67 & 10.23 & 15.34 & 20.46 & 35.46 \\
        (c) 0.5 J, 200 MeV, air & 6.14 & 8.19 & 12.28 & 16.37 & 31.37 \\
        (d) 0.5 J, 250 MeV, air & 7.67 & 10.23 & 15.34 & 20.46 & 35.46 \\
        (e) 0.5 J, 300 MeV, air & 9.20 & 12.27 & 18.40 & 24.54 & 39.54 \\
        (f) 0.5 J, 250 MeV, He+N$_2$ & 7.67 & 10.23 & 15.34 & 20.46 & 35.46 \\
        \bottomrule
    \end{tabular}
    \caption{Beamline parameters: focal distances $f_1$ and $f_2$ and positions $z_1$ and $z_2$ of the quadrupoles, the positions of the collimator $z_\mathrm{coll}$ for different beams from Fig.~\ref{fig:spectra}.}
    \label{tab:beamline_parameters}
\end{table}

Assuming that $G_1 = G_2$, the optimization criterion $f_1 = (4/3)f_2$ requires that $L_1 = (4/3)L_2$, which is why quadrupoles are taken to have the lengths of $L_1=\SI{2}{\cm}$ and $L_2=\SI{1.5}{\cm}$.
The focal lengths $f_{1,2}$ and correspondingly the positions $z_{1,2}$ of quadrupoles depend on the target energy $E_\mathrm{kin}$ of the beam and are shown in Table~\ref{tab:beamline_parameters}.

\subsection*{GEANT4 simulations}

Monte Carlo-based GEANT4 toolkit is used to simulate 3D dose deposition.
GEANT4 provides various physics lists to simulate the passage of particles. In our studies, we use the G4EmStandardPhysics\_option4 physics list for dose calculations, as it includes all physical processes involved during electron interaction with matter in the required energy range \cite{allison2016recent}. This physics list is widely recommended and used by medical physicists for dose deposition calculations  \cite{arce2025results}.
The beam data from the end of beamline simulations is saved to the disk and read by GEANT4 simulations to initialize the electron distribution.

To quantitatively assess the dose distribution characteristics in both homogeneous and heterogeneous media for the polychromatic nature of beams, we use two phantom setups. For homogeneous media testing, we employ a rectangular water phantom (\qtyproduct{30x30x30}{cm}) with a density of \SI{1}{g/cm^3}. The phantom was divided into \numproduct{500x500x300} voxels (with the last $z$ direction corresponding to beam propagation) to enable precise dosimetric analysis. For multi-field irradiation, the phantom is divided into \numproduct{300x300x300} voxels. For heterogeneous media testing, we use a water phantom of a \qtyproduct{10x10x30}{cm} size with a 2-cm-thick G4\_BONE\_COMPACT\_ICRU insert with a density of \SI{1.85}{g/cm^3} centered at \SI{3}{cm} below the phantom surface. The phantom is divided into \numproduct{300x300x300} voxels.

\section*{Acknowledgements}

The work was funded by the European Innovation Council (\textit{ebeam4theraphy} project).

\section*{Author contributions statement}

R.K. and A.G. conducted the research and wrote the paper under the supervision of A.C. and V.M. and input from other co-authors.
R.K. conducted Geant4 simulations and analysis.
A.G. conducted PIC and beamline simulations and analysis.
A.C., T.F., E.K., L.G. contributed to the beamline design.

\section*{Data availability statement}

The data is available upon reasonable request to the authors.

\section*{Competing interests}

The authors declare no competing interests.

\bibliography{sample}

\end{document}